\documentstyle[12pt]{article}
\newcommand{\rf}[1]{(\ref{#1})}
\newcommand{\beq}{\begin{equation}}
\newcommand{\eeq}{\end{equation}}
\newcommand{\g}{\gamma}
\renewcommand{\b}{\beta}
\renewcommand{\a}{\alpha}

\newcommand{\bea}{\begin{eqnarray}}
\newcommand{\eea}{\end{eqnarray}}

\newcommand{\cS}{{\cal S}}
\newcommand{\cM}{{\cal M}}

\newcommand{\cN}{{\cal N}}

\newcommand{\cU}{{\cal U}}
\newcommand{\cG}{{\cal G}}
\newcommand{\noi}{\noindent}

\newcommand{\mcg}{mapping class group }

\newcommand{\p}{{\cal P}}
\newcommand{\TR}{{\rm Tr}} 
\renewcommand{\c}{\chi_p} 
\newcommand{\MC}{\mbox{$M_{1,1}$}} 
\renewcommand{\o}{\omega}
\renewcommand{\i}{^{-1}} 
 
\newcommand{\FSS}{Fuchs, Schellekens and Schweigert }
\renewcommand{\v}{{\cal V}}
\newcommand{\N}{{\cal N}}
\newcommand{\bl}{genus one holomorphic one-point block}
\newcommand{\ver}{\thinspace\vert\thinspace}

\begin{document}
\topmargin 0pt
\oddsidemargin 5mm
\headheight 0pt
\topskip 0mm
\addtolength{\baselineskip}{0.20\baselineskip}
\pagestyle{empty}
\hfill 
\begin{center}
\vspace{3 truecm}
{\Large \bf Simple Current Extensions and Mapping Class Group 
Representations}
\vspace{5 truecm}

{\large Peter Bantay}
\vspace{1 truecm}

{\em Mathematics Department\\
University of California at Santa Cruz \\}

\vspace{3 truecm}
\end{center}
\noi
\underbar{\bf Abstract} The conjecture of  \FSS  on the 
relation of mapping class group representations and fixed point
resolution in simple current extensions is investigated, and a cohomological
interpretation of the untwisted stabilizer is given.
\vfill
\newpage
\pagestyle{plain}

\section{Introduction}

In a recent paper \cite{fss}, \FSS presented an Ansatz to describe the
modular properties of a CFT obtained by simple current extensions ( for
a review see e.g. \cite{sc} ). The crucial step in this program is the
understanding of fixed point resolution, for which they had to
introduce the notion of {\it untwisted stabilizer}. Their Ansatz
contains a set of matrices $S^J$ associated  to each simple current $J$
of the extension, which should satisfy a number of non-trivial
consistency conditions in order for the Ansatz to make sense. They have
conjectured that these consistency conditions may be satisfied by
choosing the quantities $S^J$ to be the matrices that describe the
transformation properties of the \bl s of the simple current $J$ under
the transformation $S$ that exchanges the standard homology cycles of
the torus.

The aim of this paper is to investigate the above questions in a general
setting. We will show that starting from the $S$-matrix for one-point
blocks one may define quantities that have very similar properties to
those postulated by \FSS. Moreover, we will get extra relations between
these quantities exploiting their relationship to mapping class group
representations and the extended fusion rule algebra \cite{mcg}. Our results
provide convincing evidence for the original conjecture that the
quantities appearing in the FSS Ansatz are indeed related to the
modular transformations of the one-point holomorphic blocks.

\section {The module of one-point blocks and the \\ 
generalized fusion algebra}

For each primary field $p$ let $\v(p)$ denote the space of \bl s
of $p$. This space admits a finer decomposition 
\beq \v(p)=\bigoplus_q\v_q(p),\label{vdecomp}\eeq where the subspace
$\v_q(p)$ consists of the blocks with the primary field $q$ in the
intermediate channel, and consequently its dimension is \beq
\dim\v_q(p)=N_{pq}^q.\label{fdecomp}\eeq We introduce the notation
$\p_q(p)$ for the operator projecting $\v(p)$ onto $\v_q(p)$.

It was shown in \cite{mcg} that each $\v(p)$ affords a representation
of the fusion algebra, i.e. there exist operators $\N_q(p) :
\v(p)\to\v(p)$ satisfying
\beq \N_q(p)\N_r(p)=\sum_sN_{qr}^s\N_s(p).\label{gfus}\eeq 
Moreover, it may
be shown that 
\beq \TR\left(\N_q(p)\right)=\sum_rN_{pr}^r{S_{qr}\over S_{0r}},\label{fustr}\eeq 
and that the operator $S(p) : \v(p)\to\v(p)$
implementing the modular transformation $S$ on the space of one-point
blocks diagonalizes simultaneously these generalized fusion rule
matrices, i.e.  
\beq S\i(p)\N_q(p)S(p)=\sum_r{S_{qr}\over S_{0r}}\p_r(p).\label{gverl}\eeq 

Eq. \rf{gverl} is the generalization of
Verlinde's theorem to the space of one-point blocks.
If $\a$ is a simple current, i.e. $S_{0\a}=S_{00}$, then $\N_\a(p)$ is
invertible, and one has
\beq \N_\a(p)\i \p_q(p)\N_\a(p)=\p_{\a q}(p),\label{scact}\eeq
so $\N_\a(p)$ maps $\v_q(p)$ onto $\v_{\a q}(p)$.

The operators $S(p)$ and 
\beq T(p)=\kappa\sum_q\o_q\p_q(p),\label{Tdef}\eeq 
where $\o_q=\exp(2\pi\imath\Delta_q)$ denotes the exponentiated
conformal weight - or statistics phase - of the field $q$, and 
$\kappa=\exp(-\imath\pi c/12)$ is the exponentiated central charge of the theory, 
form a representation of the \mcg $\MC$
of the one-holed torus, i.e. they satisfy the relation
\beq S(p)T(p)S(p)=T\i(p)S(p)T\i(p).\label{mod}\eeq Moreover,
\beq S^4(p)=\o_p\i{\bf 1},\label{s^4}\eeq which shows that this is a projective
representation of $SL(2,{\bf Z})$.

If $X\in \MC$ is any mapping class, then we shall denote by
$\TR_p\left(X\right)$ the trace of the operator representing it on
$\v(p)$, and we define the quantity $\c$ as \cite{mcg} 
\beq \c(X)=S_{0p}\sum_q\bar S_{pq}\TR_q(X).\label{chidef}\eeq

It is clear from its definition that $\c$ is a class function on $\MC$,
moreover it is normalized, i.e. 
\beq \c({\bf 1})=1. \label{chinorm}\eeq

We shall need  the following properties of $\c$  in the sequel.
\begin{enumerate}
\item Invariance under simple currents : if $\a$  is a simple current, then
          
	  \beq \chi_{\a p}\left(X\right)=\c\left(X\right).\label{z^1}\eeq
	  
\item Cyclicity : 

          \beq \c\left(\p_q S\p_r S\i\right)=\chi_q\left(\p_r S\p_p
	  S\i\right). \label{cyc}\eeq

\end{enumerate}

We shall also use the explicit expression of $\c\left(X\right)$ for some
mapping classes $X\in\MC$, as given in \cite{mcg}.

\section{ The FSS Ansatz}

In \cite{fss} \FSS have introduced, for any group $\cal G$ of integral
spin simple currents, a set of unitary matrices $S^\a$ for each $\a\in
{\cal G}$, whose index set consists of the fixed points of the simple
current $\a$. These matrices have to satify the following postulates :

\begin{enumerate}

\item \beq \left(S^\a T^\a\right)^3=\left(S^\a\right)^2\label{amod}\eeq
where $T^\a$ is a diagonal matrix with entries 
$T^\a_{pp}=\kappa\o_p$ for the fixed points
$p$ of $\a$.

\item
\beq \left(S^\a\right)^2_{pq}=\eta^\a_p\delta_{p,\bar q}\label{as^2}\eeq
for some phases $\eta^\a_p$.

\item
\beq \eta^\a_p\eta^\b_p=\eta^{\a\b}_p\label{etamult}\eeq

for $\a,\b\in U_p$, where the {\it untwisted stabilizer} $U_p$ is to be
defined below.

\item
\beq \eta^\a_{\bar p}=\bar\eta^\a_p\label{etabar}\eeq

\item
\beq S^\a_{p,q}=S^{\a\i}_{q,p}\label{sbar}\eeq

\item \beq S^\a_{\b p,q}=F(p,\b,\a)\theta(q,\b)S^\a_{p,q},\label{Fdef}\eeq
for some ( unknown ) phases $F$, where 
\beq \theta(q,\b)={\o_{\b q}\over\o_\b\o_q}=exp\left(-2\pi\imath Q_\b(q)\right)\label{thetadef}\eeq 
is the exponentiated monodromy charge of the primary field $q$ with 
respect to the simple current $\b$.

\item
\beq F(p,\a,\b_1)F(p,\a,\b_2)=F(p,\a,\b_1\b_2)\label{Fmult}\eeq

\item
\beq F(p,\b,\a)=F(p,\a,\b\i)\label{Fbar}\eeq

\end{enumerate}
The last two conditions on the phases $F$ show that the subset
\beq U_p=\{\a\in {\cal S}_p\ver F(p,\a,\b)=1\quad \forall\b\in\cS_p\},
\label{untwdef}\eeq 
where $\cS_p:=\{\a\ver\a p=p\}$ is the stabilizer of $p$, is actually
a subgroup of $\cS_p$ whose index is a perfect square. \FSS coined the
term {\it untwisted stabilizer} for $U_p$, and showed that it is $U_p$,
rather than $\cS_p$, that governs the fixed point resolution. They
proceeded on to show, that given a set of matrices $S^\a$ satisfying the
above conditions, one can construct a new $S$-matrix for the resolved
theory that meets the usual requirements - i.e. it is symmetric,
unitary, generates an $SL(2,{\bf Z})$ representation with an appropriate
$T$-matrix, its square is a permutation matrix -, with the possible
exception of giving integer fusion rule coefficients through Verlinde's
formula. Clearly, the question is whether there exists some natural
solutions to the above conditions, and if so whether they lead to a
consistent resolved CFT, in particular do they yield integral fusion
rule  coefficients.

Eqs. \rf{amod} and \rf{as^2} suggest that $S^\a$ and $T^\a$ might have
some relation to mapping class group representations. This idea is
supported by the observation that in the decomposition (1) of the
space $\v(\a)$, only the fixed points of the simple current $\a$ appear,
as a result of

\beq N_{\a, q}^q=\delta_{q,\a q}\label{fixa}\eeq

Moreover, the spaces $\v_q(\a)$ are all one dimensional, i.e. the space
$\v(\a)$ indeed admits a basis labeled by the fixed points of $\a$.
In such a basis the matrix $T(\a)$ representing the transformation $T\in
M_{1,1}$ would equal the matrix $T^\a$ of condition 1. above, and
this suggests that in that basis the matrix of $S(\a)$ would equal
$S^\a$, because they satisfy similar conditions. It is this argument
that led \FSS to conjecture the equality of these two quantities, i.e. that
\beq S^\a_{pq}=S(\a)_{pq}\label{FSS}\eeq

The problem is, that to make Eq.\rf{FSS} meaningful, one has to
specify a preferred  basis in the spaces $\v_q(\a)$ in order to define
the matrix elements $S(\a)_{pq}$. At first sight it is not obvious how
one should do that, but it turns at that there is indeed a canonical
basis choice, up to some trivial indeterminacy that does not show up in
the FSS Ansatz. The next section is devoted to a discussion of this
canonical basis choice and a proof of the FSS conjecture that the
resulting matrices $S(\a)_{pq}$ statisfy the requirements 1-8 above.
In order to achieve this we shall introduce quantities related to the
mapping class group action, and show that they have the same properties
as their counterparts in the FSS Ansatz.

\section{Simple current one-point blocks : \\
the canonical basis choice  and the \mcg action}

To begin with, let's recall the decomposition \beq
\v(\a)=\bigoplus_{p}\v_p(\a)\label{scdecomp}\eeq of the space of genus
1 holomorphic one-point blocks of the simple current $\a$, where $\dim
\v_p(\a)=1$ if  $\a\in\cS_p=\{\a\ver \a p=p\}$, and is $0$ otherwise.
We choose for each $\a\in\cS_p$ a vector $e_p(\a) \in \v_p(\a)$ .

Let's consider the quantity 
\beq \phi_p(\a,\b):=\TR_\a({\cal N}_\b\p_p).\label{phitrace}\eeq

From Eq. \rf{scact} we know, that $\N_\b(\a)$ maps $\v_p(\a)$ onto
$\v_{\b p}(\a)$. This implies that $\phi_p(\a,\b)=0$ if $p$ is not
fixed by both $\a$ and $\b$, while for $\a,\b\in \cS_p$ this means that 
\beq \N_\b(\a)e_p(\a)=\phi_p(\a,\b)e_{p}(\a),\label{phidef}\eeq
 and combining this with Eq.\rf{gfus} we get at once that \beq
\phi_p(\a,\b\g)=\phi_p(\a,\b)\phi_{p}(\a,\g).\label{phicoc}\eeq

Some other properties of $\phi_p(\a,\b)$ follow from Eqs. \rf{z^1} and 
\rf{cyc} if we rewrite Eq. \rf{phitrace} as 
\beq \phi_p(\a,\b)=\sum_{q,r}\theta(q,\a)\theta(r,\b)\chi_p(\p_q S\p_r S\i),
\label{phichi}\eeq
For example, we have for arbitrary simple currents $\a,\b,\g$ 
\beq \phi_p(\b,\a)=\phi_p(\a,\b\i)=\bar\phi_p(\a,\b)\label{phibar}\eeq
\beq \phi_{\g p}(\a,\b)=\phi_p(\a,\b)\label{phiz^1}\eeq
and the curious relation
\beq \sum_p\theta(p,\a)\phi_p(\b,\g)=\sum_p\theta(p,\b)\phi_p(\g,\a)\label{phicyc}\eeq

Eq.\rf{phicoc} allows us to conclude that the subset 
\beq\cU_p:=\{\a\in\cS_p\ver \phi_p(\a,\b)=1 
\qquad\forall \b\in\cS_p\}\label{Udef}\eeq 
is actually a subgroup of $\cS_p$, while as a consequence
of \rf{phibar}, the index $[\cS_p:\cU_p]$  should be a perfect square.
Note that
\beq \ver\cU_p\ver=\frac{1}{\ver\cS_p\ver}
\sum_{\a,\b}\phi_p(\a,\b).\label{Uorder}\eeq

Form Eq.\rf{gverl} we know that
\beq \cN_\b(\a)S(\a)=\sum_q\theta(q,\b)S(\a){\cal P}_q(\a),\label{NS}\eeq
and taking matrix elements of both sides in the basis $e_p(\a)$ leads to
\beq \phi_p(\a,\b)S_{pq}(\a)=\theta(q,\b)S_{pq}(\a),\label{selrule}\eeq
for $\a,\b\in \cS_p$. For a given $p$ this implies in particular
that $S_{pq}(\a)=0$ for all $q$-s having 0 monodromy charge
with respect to $\cG$ whenever there exists a $\b\in \cS_p$ with
$\phi_p(\a,\b)\ne 1$, i.e. if $\a$ does not belong to $\cU_p$. In other
words, only those $S(\a)$-s will have non-vanishing matrix elements
between $p$ and $q$-s in $I_0^\cG:=\{q\ver\theta(q,\b)=1\quad \forall
\b\in\cG\}$ for which $\a\in \cU_p$. This observation explains the role
of the subgroup $\cU_p$, which is of course nothing but the untwisted
stabilizer of \FSS.

To fully understand the relevance of $\phi_p$ and $\cU_p$, we have to
take a look now at the canonical basis choice. 
For each $p$, the space $\bigoplus_{\a\in\cS_p}\v(\a)$ admits a
binary bilinear associative operation $\Join$, such that 
the image of $\v_p(\a)\otimes\v_p(\b)$ lies in 
$\v_p(\a\b)$, i.e.
\beq e_p(\a)\Join e_p(\b)=\vartheta_p(\a,\b)e_p(\a\b),\label{opdef}
\eeq where $\vartheta_p(\a,\b)$ is some 2-cocycle of the stabilizer $\cS_p$.
In general $\Join$ is not commutative, because we have
\beq e_p(\b)\Join e_p(\a)=\phi_p(\a,\b)e_p(\a)\Join e_p(\b),\label{com}\eeq
which implies at once that
\beq\vartheta_p(\b,\a)=\phi_p(\a,\b)\vartheta_p(\a,\b),\label{phicom}\eeq
i.e. $\phi_p$ is the so-called commutator cocycle of $\vartheta_p$. 

It is a well-known result of group cohomology, that the cocycle
$\vartheta_p$ is trivial if and only if $\phi_p(\a,\b)=1$ for all
$\a,\b$. But this condition holds for $\cU_p$ by definition,
consequently there exists some function $\zeta_p$ on $\cU_p$ such that
\beq\vartheta_p(\a,\b)=\frac{\zeta_p(\a\b)}{\zeta_p(\a)\zeta_p(\b)}
\label{cob}\eeq
for all $\a,\b\in \cU_p$. But then the rescaled basis vectors
\beq \hat e_p(\a)=\zeta_p(\a) e_p(\a)\label{canbas}\eeq
will satisfy
\beq \hat e_p(\a)\Join \hat e_p(\b)=\hat e_p(\a\b) \label{repr}\eeq
for $\a,\b \in \cU_p$.

It is this last condition, Eq.\rf{repr} that fixes our canonical basis
choice, i.e. the basis consisting of the vectors $\hat e_p(\a)$. 
Of course, there is still some indeterminacy left, because we are still
free to rescale the vectors by
\beq \hat e_p(\a)\mapsto \psi(\a)\hat e_p(\a),\label{gauge}\eeq
where $\psi(\a)$ is some 1-dimensional character of the group $\cU_p$,
so that the resulting basis will still satisfy Eq.\rf{repr},
but it is obvious that this indeterminacy does not show up in the FSS
Ansatz.

Let's now take a look at
\beq \eta(p,\a):=\TR_\a(S^2\p_p)=\sum_q\theta(q,\a)\chi_q(S^2\p_p)\label{etadef}\eeq
which obviously vanishes if $p$ is not self-conjugate or not fixed by
$\a$, while for a self-conjugate fixed point its value is
$\eta(p,\a)=S^2(\a)_{pp}$. 
Moreover, it follows from Eq. \rf{etadef} that $\eta(p,\a)=0$ if $\a$ has
non-trivial monodromy with respect to any simple current, and that
\beq \eta(\b p,\a)=\eta(\bar p,\a)=\eta(p,\a).\label{etaz^1}\eeq
The relation of $S(\a)^2$ to $\Join$ gives at once that
\beq  \eta(p,\a\b)=\eta(p,\a)\eta(p,\b)\qquad{\rm for}\quad
\a,\b\in\cU_p.\label{??} \eeq

From the results of \cite{mcg} we can actually
compute explicitly $\eta(p,\a)$, leading to
\beq \eta(p,\a)=\nu_p\sum_{q,r}\theta(q,\a)N_{qr}^pS_{0q}S_{0r}
{\o_q^2\over\o_r^2},\label{etaexpr}\eeq
where $\nu_p$ is the Frobenius-Schur indicator \cite{FS} of the field $p$,
which is 0 if $p$ is not self-conjugate, and is $\pm 1$ according to 
whether $p$ is real or pseudo-real. 

The next quantity we introduce is
\beq \mu(p,\a):=\kappa^3\o_p\TR_\a(S\p_p)=\kappa^3\o_p S_{pp}(\a).\label{mudef}\eeq

First, we rewrite Eq. \rf{mudef} in the form
\beq \mu(p,\a)=\o_p\i\sum_{q,r}\o_q\i\theta(r,\a)\chi_r(\p_p S\p_q S\i),\label{muchi}\eeq
which in conjunction with Eq.  \rf{z^1} leads to
\beq \mu(\b p,\a)=\o_\b\i\theta(p,\b)\i\mu(p,\a),\label{muz^1}\eeq
and
\beq \mu({\bar p},\a)=\mu(p,{\bar \a})=\mu(p,\a).\label{mubar}\eeq

If we introduce the quantities
\beq \cM_k(\a,\b):=\sum_p\o_p^{1-k}\theta(p,\b)\mu(p,\a),\label{Mdef}\eeq
then a  simple argument involving Eq. \rf{muz^1} shows that
\beq \cM_k(\a,\b\g^k)=\theta(\b,\g)\o_\g^k\cM_k(\a,\b),\label{Mtrans}\eeq
and 
\beq \cM_k(\a,\a\b)=\o_\a\cM_k(\a,\b).\label{Mtrans2}\eeq

The cyclicity property Eq. \rf{cyc}, together with Eqs. \rf{phichi},
\rf{etadef} and \rf{muchi}, gives the sum rules
\bea\cM_0(\a,\b)&=&\sum_p\o_p\i\phi_p(\a,\b), \label{M0}\\
    \cM_4(\a,0)&=&\kappa^6\sum_p\o_p\eta(p,\a),\label{M4}\eea
which connect $\mu(p,\a)$ with $\phi_p(\a,\b)$ and $\eta(p,\a)$,
while Eq. \rf{Mtrans}  and the explicit trace formulae of \cite{mcg} give
\bea\cM_1(\a,\b)&=&\kappa^9\o_\b\sum_{p,q,r}N_{pq}^rS_{0p}S_{0q}S_{0r}
\theta(p,\a){\o_p^6\o_r^3\over \o_q^2},\label{M1}\\\nonumber\\ 
\cM_2(\a,\b^2)&=&\kappa^9\o_\b^2\sum_{p,q,r}N_{pq}^rS_{0p}S_{0q}S_{0r}
\theta(p,\a){\o_p^4\o_r^4\over \o_q^2}.\label{M2}\eea

\section{ Conclusions}

In the previous section we have investigated the structure of the space
of genus one holomorphic one-point blocks of the simple currents, and
have defined various quantities - such as $\phi$ and $\eta$ - that
characterize the action of the \mcg on these spaces. We have seen that
these quantities obey a host of non-trivial relations. We have also been
able to specify a canonical basis choice - up to some trivial
indeterminacy - in the space of one-point blocks, allowing us to give an
invariant meaning to the matrix elements of the operators representing
the mapping classes. Finally, we have introduced in an invariant way the
subgroup $\cU_p$ of the stabilizer $\cS_p$, and gave a cohomological
interpretation of its origin, related to the canonical basis choice.

Comparing our results with the postulates $1,\ldots, 8$ of Section 3,
the truth of the FSS conjecture is obvious, since upon identifying 
$S^\a$ with $S(\a)$ - with respect to the canonical basis! -, 
$\phi_p(\a,\b)$ is identified with $F(p,\a,\b)$, $\cU_p$ with $U_p$,
and $\eta(p,\a)$ with $\eta_p^\a$, and all of the conditions on these
quantities are indeed fulfilled. 

Unfortunately, the implementation of the canonical basis choice, as
well as the explicit computation of the matrices $S(\a)$, is not a
straightforward matter in general. To our best knowledge it had only
been done in two class of models up to now, namely WZNW models through
the use of orbit Lie-algebras \cite{orblie}, and holomorphic orbifold
models through the application of the theory developped in
\cite{orbmcg}. It had been verified in these two cases by explicit
numerical checks, that the FSS Ansatz does not only give a correct
$SL(2,{\bf Z})$ representation, but indeed the one of the resolved
theory.
\vspace{1.5 truecm}

{\it Acknowledgement :} It is a pleasure to acknowledge discussions with
Geoff Mason, Christoph Schweigert and Peter Vecserny\'{e}s.

\vspace{2 truecm}

Supported partially by OTKA T016251

e-mail: bantay@hal9000.elte.hu

\end{document}